\documentclass[twocolumn,times,trackchanges]{aastex63}

\usepackage{xspace}

\newcommand{\teff}{\ensuremath{T_{\mathrm{eff}}}\xspace}

\newcommand{\msun}{$M_\odot$\xspace}
\newcommand{\rsun}{$R_\odot$\xspace}

\newcommand{\rearth}{$R_\oplus$\xspace}
\newcommand{\snr}{$\mathrm{SNR_1}$\xspace}

\newcommand{\sesn}{$\theta_{\mathrm{SE/SN}}$\xspace}

\received{\today}
\submitjournal{AAS Journals}

\shorttitle{A Rising Fraction of Super-Earths with Age}
\shortauthors{Sandoval et al.}
\graphicspath{{./}{figures/}}

\begin{document}


\title{The Influence of Age on the Relative Frequency of Super-Earths and Sub-Neptunes}

\correspondingauthor{Angeli Sandoval}
\email{angelis0821@gmail.com}

\author[0000-0003-1133-1027]{Angeli Sandoval}
\affiliation{Department of Physics and Astronomy, Hunter College, City University of New York, New York, NY 10065, USA}

\author[0000-0002-3011-4784]{Gabriella Contardo}
\affil{Center for Computational Astrophysics, Flatiron Institute, New York, NY 10010, USA}

\author[0000-0001-6534-6246]{Trevor J.\ David}
\affil{Center for Computational Astrophysics, Flatiron Institute, New York, NY 10010, USA}
\affil{American Museum of Natural History, Central Park West, New York, NY 10024, USA}

\begin{abstract}
There is growing evidence that the population of close-in planets discovered by the Kepler mission was sculpted by atmospheric loss, though the typical timescale for this evolution is not well-constrained. Among a highly complete sample of planet hosts of varying ages the age-dependence of the relative fraction of super-Earth and sub-Neptune detections can be used to constrain the rate at which some small planets lose their atmospheres. Using the California-Kepler Survey (CKS) sample, we find evidence that the ratio of super-Earth to sub-Neptune detections rises monotonically from 1--10~Gyr. Our results are in good agreement with an independent study focused on stars hotter than the Sun, as well as with forward modeling simulations incorporating the effects of photoevaporation and a CKS-like selection function. We find the observed trend persists even after accounting for the effects of completeness or correlations between age and other fundamental parameters.
\end{abstract}

\keywords{Exoplanets (498) --- 
Exoplanet evolution (491) ---
Exoplanet astronomy (486) ---
Super Earths (1655) --- 
Mini Neptunes (1063)}

\section{Introduction} \label{sec:intro}
NASA's Kepler mission led to the discovery of thousands of small exoplanets, most of which have  sizes between 1 and 4 Earth radii and orbital separations smaller than that of Mercury \citep{Borucki2010}. The typical Kepler planet thus differs greatly in its size and placement from what we have come to know from the planets in our Solar System. Subsequent planet mass measurements from radial velocities and transit-timing variations revealed a diversity of compositions for small planets, even among planets within the same system. In one notable case the two planets of Kepler-36 differ by an order of magnitude in bulk density despite having orbital separations that differ by only 10\%, or 5 Earth-Moon separations in physical units \citep{Carter2012}. 

These observations led theorists to suggest that atmospheric loss might explain the differences in small planet densities. Early theoretical studies of atmospheric loss in the planet mass and orbital separation regime relevant to Kepler focused on the effects of photoevaporation by high energy stellar radiation \citep{Lopez2013, OwenWu2013}. Those works were the first to predict the existence of the radius valley, a lack of intermediate-sized sub-Neptune planets due to unstable configurations of core mass, atmospheric mass fraction, and high energy emission exposure. However, the valley could not be resolved in earlier Kepler catalogs due to large uncertainties in stellar, and thus planetary, radii.

Precise characterization of Kepler planet host stars revealed the predicted valley in the size distribution of small exoplanets \citep{Fulton2017}. Those authors found a relative scarcity of close-in ($P < 100$~days) exoplanets with sizes between 1.5 and 2.0 Earth radii in the completeness-corrected size distribution of small planets. While the photoevaporation theory predicted the existence of this feature in the close-in exoplanet population, recent theoretical work on the core-powered mass loss mechanism is also able to successfully reproduce the radius valley by means of the planet’s internal luminosity \citep{Gupta2019, Gupta2020}. In this model, the planet’s luminosity energy can be greater or equal to the planet’s atmospheric binding energy therefore causing the planet to lose atmosphere and decrease in size over time. One notable difference between the two theories is the predicted timescale. The photoevaporation process is thought to primarily occur during the first 100 Myr of a host star’s life, when the stellar high energy output is at its highest, while core-powered mass loss happens over gigayears as planetary cores slowly cool and contract. It is important to note that the 100 Myr timescale for photo-evaporation is typically attributed to Sun-like stars, which this paper covers. There is also evidence for a radius valley among planets orbiting low mass stars \citep{Wu2019, Cloutier2020, vanEylen2021} which remain active on longer timescales, more comparable to the timescale of core-powered mass loss.

Although there is compelling evidence of atmospheric loss among low-mass planets, it is still unclear what mechanism is the main driver. Determining the typical timescale for atmospheric loss may help further our understanding of the processes influencing the evolution of close-in, low-mass planets. In this paper we find that the ratio of detected super-Earths to sub-Neptunes increases on gigayear timescales, which is consistent with atmospheric loss but in tension with the timescale for photoevaporation. In \S\ref{sec:sample} we discuss the selection of our stellar and planetary samples. Our analysis steps are described in \S\ref{sec:analysis}. Finally, our primary findings and the implications for exoplanet evolution models are summarized in \S\ref{sec:conclusions}.

\section{Sample Selection} \label{sec:sample}
In this study we use the California-Kepler Survey (CKS) sample, which is a subset of Kepler planet host stars with precise, accurate, and uniform parameters derived from high-resolution spectroscopy \citep{Petigura2017, Johnson2017}. Specifically, we use the CKS VII sample \citep{Fulton2018} which presented updated stellar and planetary radii, as well as updated stellar ages, based on the inclusion of Gaia DR2 parallaxes. The ages in the CKS VII study were computed using the \texttt{isoclassify} package \citep{Huber2017} which relies on isochrones derived from MESA stellar evolution models \citep[the MIST v1.1 models,][]{Choi2016, Dotter2016}. 

Interpretation of our results relies on the accuracy of the stellar ages. In a companion paper we showed that the CKS VII ages are generally accurate and reliable \citep{David2020}. While we do not reproduce the age validation analysis, we briefly summarize the results here. Those authors found the CKS ages compare favorably to those determined by \citet{SilvaAguirre2015} from  asteroseismology as well as to gyrochronology ages derived using the relations of \citet{Angus2019}. The median log(age) uncertainty in the CKS sample is also close to the scatter from the asteroseismology and gyrochronology comparison samples, indicating that the CKS age uncertainties are estimated appropriately. Finally, the \citet{David2020} study also examined trends between CKS ages and independent age indicators. Those authors found that photometric variability amplitude and near-UV excess decline predictably with age, while the velocity dispersion increases with age. These trends confirm that the CKS ages are broadly accurate.

We performed several cuts to the CKS VII catalog to focus on small, close-in planets with reliable parameters. There are 1901 planet candidates orbiting 1189 unique host stars in the CKS VII catalog. We excluded host stars with $>4\sigma$ disagreements between the Gaia DR2 trigonometric parallaxes \citep{GaiaDR2} and isochrone-derived parallaxes from \citet{Fulton2018}, as well as stars lacking age or mass measurements. We also excluded planets with false positive dispositions in \citet{Petigura2017} and planets with periods~$>$~100 days. 

As we are measuring the ratio of super-Earth to sub-Neptune detections as a function of age, we want to ensure that the sensitivity to small planets is not a strong function of age. As in \citet{Petigura2018} and \citet{Berger2020b} we computed the single-transit signal-to-noise of a hypothetical Earth-sized planet for each star using the equation,

\begin{equation}
    \text{SNR}_1 = \left ( \frac{R_\oplus}{R_\star} \right )^2 \left ( \frac{1}{\text{CDPP3}}\right ),
\end{equation}

where CDPP3 is the combined differential photometric precision on 3-hour timescales \citep{Christiansen2012}. We observed a trend in $\mathrm{SNR_1}$ with age such that older stars have less sensitivity to small planets on average, owing to their preferentially larger radii. Over a grid of threshold values we computed Spearman's rank correlation coefficient between the ages and $\mathrm{SNR_1}$ values for subsets of the CKS sample with $\mathrm{SNR_1}$ values above the threshold. We found that the $p$-value is $\gtrsim$ 0.02 if the $\mathrm{SNR_1}$ threshold is set to 0.45. We thus excluded stars with $\mathrm{SNR_1}<0.45$ to mitigate the correlation between age and small planet sensitivity. We address the issue of completeness further in \S\ref{subsec:completeness}. 

\begin{figure}
    \centering
    \includegraphics[width=\linewidth]{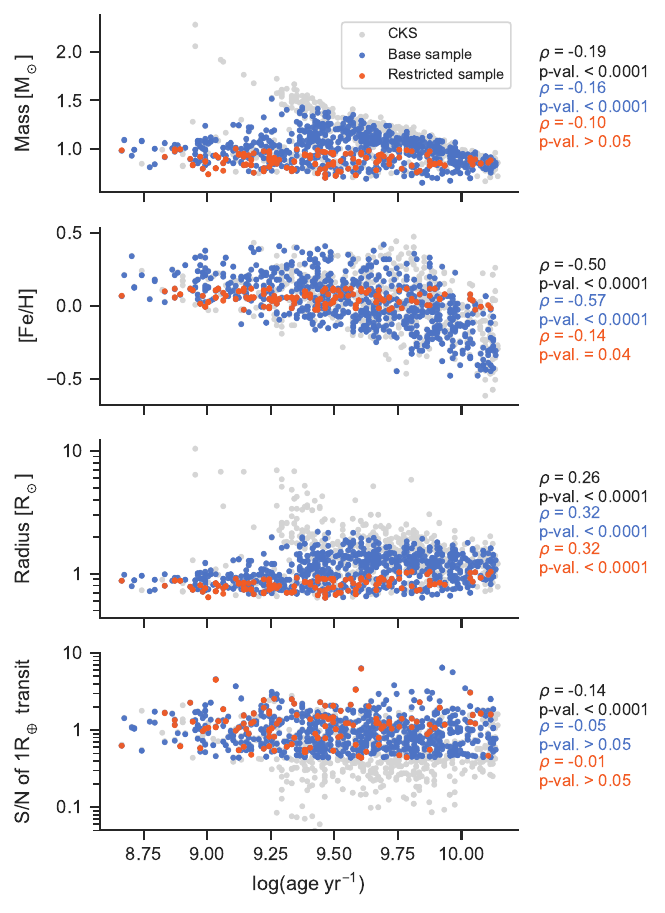}
    \caption{Correlations with stellar age in the CKS sample (grey points), our base sample (blue), and the restricted sample (red). Spearman's rank correlation coefficient ($\rho$) and $p$-values for each sample are printed to the right of each panel. The bottom panel shows the age-dependence of the signal-to-noise of a single transit by a hypothetical 1~$R_\oplus$ planet, defined as \snr in \S\ref{sec:sample}.}
    \label{fig:sample}
\end{figure}

After all of the cuts described above, we are left with a sample of 1158 planets orbiting 718 unique hosts. We refer to this as our base sample. As shown in Fig.~\ref{fig:sample}, age is correlated with stellar mass, radius, and metallicity in the base sample, making it difficult to isolate age as an independent variable in our analysis. To this end, we constructed a restricted sample with the goal of reducing correlations between age and other fundamental stellar parameters. The restricted sample was created from the following cuts: exclusion of false positives, $P<100$~days, $R_P<5$~\rearth, $\mathrm{SNR_1}>0.45$, $M_\star<1$~\msun, $R_\star<1.05$~\rsun, and [Fe/H] in the range [-0.025,0.125] dex, i.e. within 0.075 dex of the median metallicity in the sample. Bounds on stellar mass, radius, and metallicity were chosen in a manner similar to the \snr threshold, described above, while also preserving a sizable sample of 189 planets orbiting 112 unique hosts. Though the restricted sample is not completely devoid of biases, correlations between age and other parameters are made significantly weaker by the restrictions as indicated in Fig.~\ref{fig:sample}.

\section{Analysis} \label{sec:analysis}

\subsection{Completeness}
\label{subsec:completeness}
\begin{figure*}
    \centering
    \includegraphics{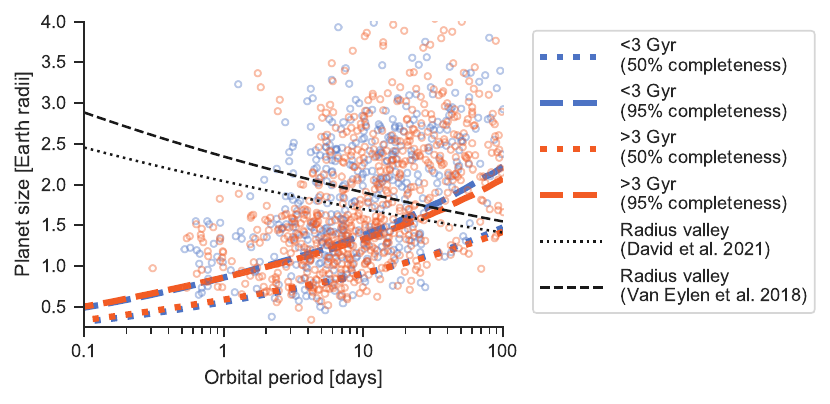}
    \caption{Distribution of planets younger than 3~Gyr (blue points) and older than 3~Gyr (red points) in the period-radius diagram for our base sample. The thick dotted/dashed blue and red lines indicate the mean 50\%/95\% completeness contours for the host stars in the young and old samples. The black dashed and dotted lines show two equations derived for the radius valley.}
    \label{fig:completeness}
\end{figure*}

The Kepler time series for each star differs in its sensitivity to planets of a given size and period due to differences in, e.g., stellar size, brightness, and activity. The results of our study would be impacted if there is a sizable systematic difference in the pipeline completeness between the younger and older stars in our sample. To assess whether pipeline completeness changes significantly as a function of age in the CKS sample we used the methodology of \citet{Burke2015} to construct average completeness maps specifically for CKS host stars in different age bins.\footnote{The code at \url{https://dfm.io/posts/exopop/} was adapted for this purpose.} We used the stellar masses and radii published in the CKS VII catalog for this exercise. We found that the mean completeness, or detection efficiency, does not change substantially for our base sample across all ages. In fact, there is a modest enhancement in detection efficiency at all periods among stars younger than 1~Gyr relative to stars older than 1~Gyr likely due to the preferentially larger radii of the older stars. As we discuss in the following section, the lower average completeness at older ages would serve to strengthen the observed trend in the ratio of super-Earth to sub-Neptune detections, if properly accounted for. As shown in Fig.~\ref{fig:completeness}, the mean completeness contours for stars younger than 3~Gyr and older than 3~Gyr in our base sample are nearly identical, differing by no more than 7\% in the period range of 0.1-100 days. The similarity in completeness between young and old stars in our sample is likely due to the \snr cut described in \S\ref{sec:sample}. We conclude that completeness corrections are not likely to substantially impact the results of our analysis described in the following section. 

\subsection{Monte Carlo simulation}
The core of our analysis is a Monte Carlo simulation measuring the relative number of super-Earth and sub-Neptune detections as a function of stellar age. For convenience, we define the ratio of super-Earth to sub-Neptune detections within a given sample of stars as the parameter \sesn. In each of 5000 simulations, we model planet radius and stellar age uncertainties as normal distributions. Orbital period uncertainties from Kepler are negligible and not modeled here. In the CKS VII sample the upper and lower planet radius uncertainties are symmetric. The age uncertainties, however, are not symmetric, and the upper uncertainties on log(age) are smaller than the lower uncertainties by approximately 0.1 dex on average. As the actual age posteriors are not available, for each star we instead model the age as a normal distribution (in logarithmic space) centered on the median posterior age with a width equivalent to the maximum of the upper and lower uncertainties. We also explored sampling the ages from uniform distributions but found that the normal distribution approximation provides a better match to the empirical age CDFs. 

After simulating planetary radii and ages as described above, the ratio of super-Earth to sub-Neptune detections is then computed in 50 evenly-spaced, overlapping log(age) bins of width 0.5 dex. The bin centers range from log(age) = 8.25 to 10 dex. To separate planets into super-Earths and sub-Neptunes we used three equations of the radius valley: the period-dependent equations derived by \citet{vanEylen2018} and \citet{David2020} as well as a valley at 1.8~\rearth that is flat with orbital period. We recover the same qualitative trend with age regardless of the radius valley prescription. Furthermore the statistical significance of the trend does not change drastically by adopting different radius valley equations. We  also explored incorporating the reported statistical uncertainties on the slope and intercept of the radius valley. However, modeling the slope and intercept as Gaussian and independent is a poor assumption and results in unreliable results. We ultimately adopted the radius valley equation presented in \citet{David2020}.

\begin{figure*}
    \centering
    \includegraphics[width=0.45\linewidth]{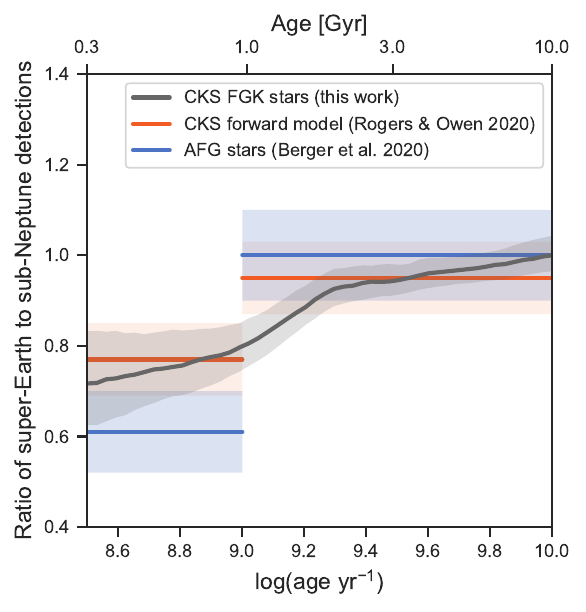}
    \includegraphics[width=0.45\linewidth]{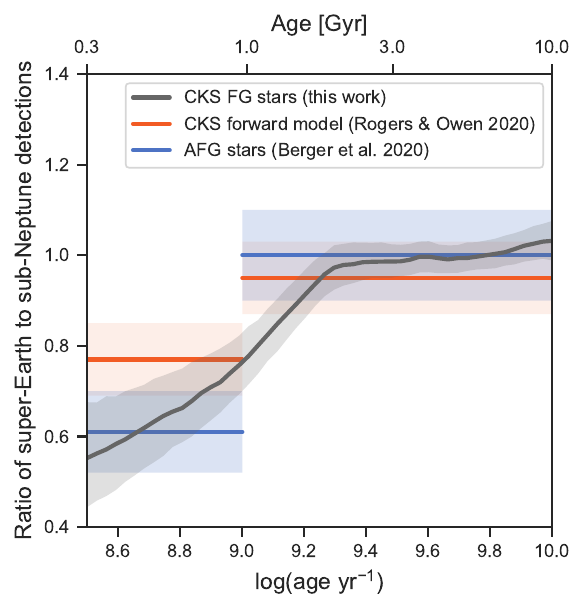}
    \caption{Results of the Monte Carlo simulation described in \S\ref{sec:analysis}. Shaded regions indicate $1\sigma$ uncertainties, in the case of our study determined from the 68.3\% percentile range. \textit{Left:} results for the base sample of FGK stars. \textit{Right:} results for the base sample restricted to stars with $\teff>5500$~K.}
    \label{fig:ratio}
\end{figure*}

The results of our simulations are shown in Fig.~\ref{fig:ratio}. For the base sample, we found \sesn increases from $0.76\pm0.08$ at $<1$~Gyr to $0.94\pm0.07$ at $>$1~Gyr, where we have quoted the mean and standard deviations from the Monte Carlo simulations. As there are only 70 stars with median posterior ages less than 1~Gyr in our base sample, we also computed the same ratios using 3~Gyr as the dividing line between young and old samples. In that case we found \sesn increases from $0.82\pm0.1$ at $<3$~Gyr to $0.97\pm0.04$ at $>3$~Gyr. 

As isochronal ages tend to be more precise and reliable for hotter main sequence stars, we repeated our analysis by further restricting the base sample to stars with \teff $>5500$~K. In this case, we found a slightly more pronounced difference between the young and old planet populations: \sesn increases from $0.65\pm0.11$ at $<1$~Gyr to $0.96\pm0.08$ at $>1$~Gyr. Similarly, using  using 3~Gyr to divide young and old planet populations we found \sesn = $0.78\pm0.17$ among young planets and $1.00\pm$0.04 among old planets.

As mentioned above, the upper and lower age uncertainties from the CKS sample are not always equal, indicating that age posteriors are not generally symmetric. We therefore investigated the degree to which asymmetric probability distributions in age impact the results of the Monte Carlo simulation. For each star, we fit a skew normal point percent function (PPF) to the reported 16th, 50th, and 84th age percentiles published in the CKS catalog. We used the \texttt{scipy.stats} package to generate skew normal PPF models and the \texttt{lmfit} package to perform a least-squares fit to the reported percentiles to find best-fit shape parameters. Then, in the Monte Carlo simulation we used the best-fit shape parameters to draw ages randomly from skew normal, rather than normal, distributions. Adopting skew normal distributions to sample the ages in the simulation resulted in changes to the \sesn ratios quoted above that were $\ll 1\sigma$ in each case. The uncertainties in the \sesn ratios were similarly unchanged.

While the statistical significance of the observed trends is always $<3\sigma$, our results and uncertainties are in good agreement with previous findings in the literature, as discussed in \S\ref{sec:conclusions}. 

\subsection{Effects of stellar mass, radius, and metallicity}
Stellar age is correlated with the stellar mass, radius, and metallicity. As discussed in \S\ref{sec:sample}, stellar radius affects sensitivity to small planets. Old stars are preferentially larger and thus less sensitive to small planets. However, by performing cuts on \snr and examining completeness as a function of age we are able to rule out the possibility that the observed trend in \sesn is due to completeness effects.

\begin{figure}
    \centering
    \includegraphics[width=\linewidth]{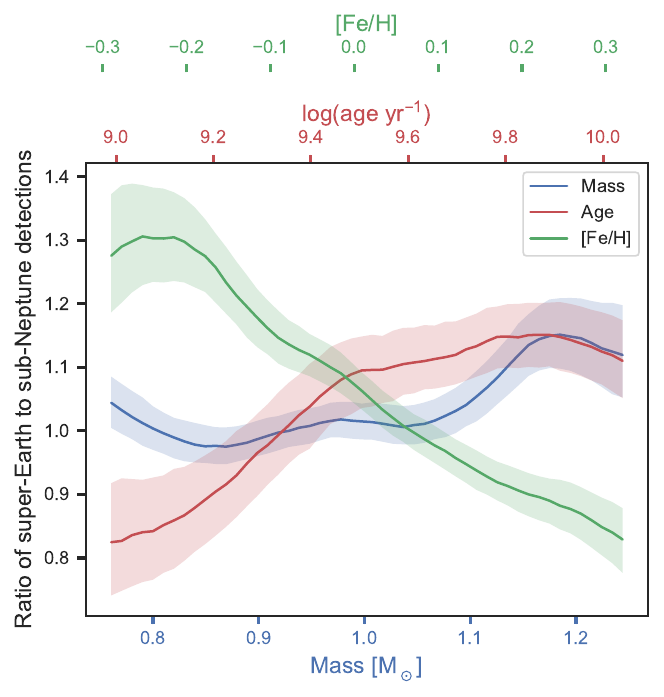}
    \caption{Trends in the ratio of super-Earth to sub-Neptune detections in the base sample as a function of stellar mass (blue), age (red), and metallicity (green).}
    \label{fig:trends}
\end{figure}

Various authors have pointed out that the radius valley and small planet size distribution also depend on the stellar mass \citep[e.g.][]{Fulton2018, Wu2019, Berger2020b} and metallicity \citep{Owen2018}. Given the correlations between these parameters and age, we investigated whether the observed trend in \sesn might be more closely linked with one of these other parameters. We repeated the Monte Carlo simulation described above for the base sample by binning in age, mass, and metallicity. In this case, computed \sesn in 50 overlapping bins from the 5th to 95th percentile values in each parameter. For the bin widths in each parameter, we chose 3$\times$ the width given by Scott's normal reference rule. The results of this exercise are shown in Fig.~\ref{fig:trends}. 

We observe that the trend in stellar mass is mostly flat and conclude that stellar mass is unlikely to be a confounding variable in our analysis. However, the \sesn parameter is sensitive to metallicity. There are more super-Earths, relative to sub-Neptunes, at lower metallicities, which are strongly correlated with older ages in the  sample. In the base sample, the \sesn parameter decreases from $1.23 \pm 0.09$ among host stars in the lower third of metallicity values ([Fe/H]$<$-0.03 dex) to $0.89 \pm 0.05$ among stars in the upper third ([Fe/H]$>$+0.11 dex). Thus, it is unclear whether the observed trend is primarily due to age or metallicity effects. We also note that the period distribution of small planets depends sensitively on metallicity, with planets found at preferentially larger periods around metal-poor stars \citep{Petigura2018}. This effect may conceivably introduce a gradient in \sesn with metallicity.

\begin{figure}
    \centering
    \includegraphics[width=\linewidth]{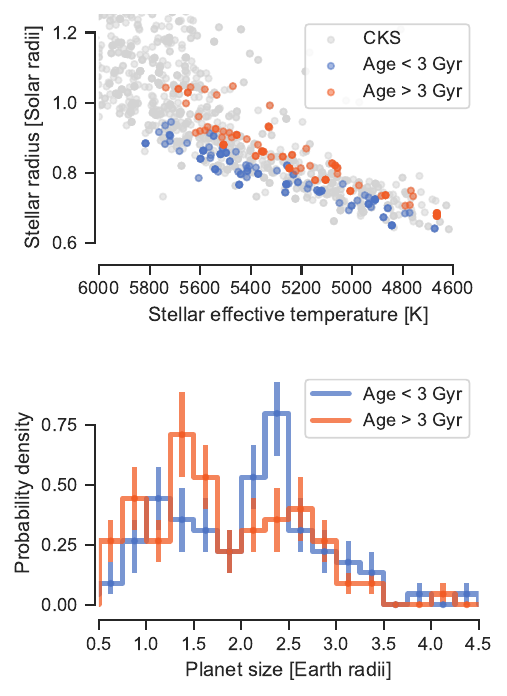}
    \caption{Comparison of H-R diagram positions (top) and planet size distributions (bottom) for the unbiased sample. Histogram errorbars were determined from bootstrapping simulations.}
    \label{fig:unbiased}
\end{figure}

To further isolate age as an independent variable, we constructed the restricted sample as discussed in \S\ref{sec:sample}. In this sample, we have effectively limited the correlation between age, mass, and metallicity as much as possible while retaining a sizable population of planets. The cuts described effectively limited our sample to cool stars (GK dwarfs) in a very narrow metallicity range around the median metallicity represented in the base sample. We then examined the radius distribution among planets with median posterior ages less than and greater than 3~Gyr. There were 112 planets and 90 planets in the young and old samples, respectively. We performed 5000 bootstrapping simulations, selecting 90 planets at random (with replacement) from each of two populations to produce the histograms and errors shown in Fig.~\ref{fig:unbiased}. We found that, even in an extremely narrow range of metallicity, the \sesn parameter increases from $0.66\pm0.14$ to $1.28\pm0.28$ between the young ($<3$~Gyr) and old ($>3$~Gyr) planet samples. We additionally performed a k-sample Anderson-Darling test on the young and old planet size distributions in this restricted sample, finding a test statistic of 1.5 and a $p$-value of 0.08, indicating the null hypothesis can be rejected at the 10\% level.

The code and data tables required to reproduce the results of this study are made publicly available.\footnote{\url{https://github.com/angelis21/Kepler-Exoplanet-Evolution}}

\section{Discussion \& Conclusions} \label{sec:conclusions}
Using the CKS catalog of Kepler planet parameters we found that the ratio of super-Earth to sub-Neptune detections increases with system age between $\sim$1--10~Gyr (Fig.~\ref{fig:ratio}). We showed that the observed trend is not due to completeness effects, and in fact the trend would only strengthen if completeness were accounted for: relative to sub-Neptunes, super-Earths appear to be more common around older stars despite the lower average completeness and increased difficulty of detecting small planets around older, larger stars.

Age is correlated with other fundamental parameters in the CKS planet host sample. The correlation between age and metallicity is particularly strong, and a previous study of the CKS sample has suggested a link between metallicity and the planet size distribution \citep{Owen2018}. However, while the limited sample size makes the complete removal of these correlations difficult, we showed that even in a restricted sample of mostly unbiased planet hosts the small planet radius distribution seems to depend sensitively on system age (Fig.~\ref{fig:unbiased}).  

Recent studies have also shown that the size distribution of small planets depends on the age of the planet population. In a companion paper, \citet{David2020} found evidence that the precise location of the radius valley changes over gigayear timescales, owing primarily to a lack of large super-Earths at ages $<$2--3~Gyr. That result is in agreement with our finding that the size distribution of close-in, low-mass planets evolves over similar timescales.

\citet{Berger2020b} also examined the size distributions of small planets for young and old planet populations and first pointed out the age trend in the ratio of super-Earth to sub-Neptune detections, using independently determined stellar parameters. Those authors found that ratio increases from 0.61 $\pm$ 0.09 at young ages ($<$1~Gyr) to 1.00 $\pm$ 0.10 at older ages ($>$1~Gyr), in agreement with our findings. A key difference between that study and the present work is the underlying temperature range of the host star samples. Those authors focused on hotter stars (5700--7900~K) for increased reliability in isochronal age estimates. Our study shows that the age-dependence of the super-Earth to sub-Neptune detection ratio is also present among planets orbiting cooler host stars ($<$5800~K), thus extending the trend to cooler temperatures and lower stellar masses for the first time.

A more direct comparison is provided by the recent work of \citet{RogersOwen2020}, who presented forward modeling simulations incorporating an analytical photoevaporation model, Kepler completeness maps, and a CKS-like selection function. Those authors predicted that the ratio of super-Earths to sub-Neptunes (where they separated the two classes at $R_P$=1.8) varies from 0.77 $\pm$ 0.08 at ages $<$1 Gyr to 0.95 $\pm$ 0.08 at ages $>$1 Gyr, again in agreement with our findings. While the statistical significance of the trend noted here is only $\approx2\sigma$ it is on par with the previously mentioned studies. A larger planetary sample and higher precision ages would help to confirm or invalidate the results presented here. 

If the evolution of Kepler planet sizes on gigayear timescales can be confirmed more theoretical work is required to interpret this result. In the photoevaporation model, though the majority of mass-loss is expected to occur within the first 0.1 Gyr, some fraction of sub-Neptunes are converted to super-Earths on longer timescales \citep{RogersOwen2020}. Furthermore, using two relations that describe the age-dependence of stellar X-ray emission and the ratio of X-ray to extreme ultra-violet emission, \citet{King2020} found that the extreme ultra-violet emission rate continues to be substantial even after the first 0.1 Gyr of a host star’s life, indicating that photoevaporation may be significant on gigayear timescales.

As pointed out by previous authors \citep[e.g.][]{Berger2020b, Cloutier2020}, many predictions of the photoevaporation and core-powered mass-loss models are so similar that current data are not able to conclusively favor one model over the other. This is true for the stellar mass dependence  and orbital period dependence of the radius valley location \citep[e.g.][]{LopezRice2018, Wu2019, Gupta2019, Gupta2020}. As discussed above, it is also no longer clear if the range of timescales for conversion of sub-Neptunes to super-Earths can be used to decisively discriminate between the photoevaporation and core-powered mass-loss theories. While our results provide further insights into the nature of sub-Neptunes and super-Earths, it may be necessary to identify  other parameters to determine which mechanisms of atmospheric loss play a substantial role in the evolution of small planets.

\acknowledgments A.S. is supported in part by the AstroCom NYC program (NSF award AST-1831412) and Simons Foundation (Award \#533845). We thank Dan Foreman-Mackey, David W. Hogg, and Erik Petigura for helpful discussions. This paper includes data collected by the {\em Kepler} mission, funded by the NASA Science Mission directorate. 

\facilities{Kepler}

\newpage


\end{document}